# Using Machine Learning to Predict Poverty Status in Costa Rican Households

Ji Yoon Kim

*Johns Hopkins University, Carey Business School*
Washington, DC
jkim624@jhu.edu

**Abstract – This study presents two supervised multiclassification machine learning models to predict the poverty status of Costa Rican households as a way to support government and business sectors make decisions in a rapidly changing social and economic environment. Using the Costa Rican household dataset collected via the proxy means test conducted by the Inter-American Development Bank, Random Forest and Gradient Boosted Trees achieved F1 scores of 64.9% and 68.4%, respectively. This study also reveals that education has the greatest impact on predicting poverty status.**

**Keywords – Poverty Prediction, Supervised Machine Learning, Multiclassification, Random Forest, Gradient Boosted Trees**

## I.    INTRODUCTION

Over the past two decades, Costa Rica has removed their foreign investment restrictive measures and liberalized their international trade policies [1]. These efforts have brought economic growth to Costa Rica and have led Costa Rica to become an upper-middle-income country. According to the World Bank Group [1], when the poverty line of upper-middle-income countries was set to $5.50 per day, Costa Ricans with incomes below the poverty line decreased, from 12.9% in 2010 to 10.6% in 2019. Despite strong economic growth, Costa Rica has recently been experiencing economic hardships due to the COVID-19 pandemic. Costa Rica's gross domestic product (GDP) decreased by 4.1% in 2020. A sharp increase in unemployment pushed an estimated 124,000 people into poverty, which raised the poverty rate to 13.0% in the same year [1].

To minimize the social and economic impacts of unexpected crises, it is necessary to consider introducing data-driven technology capable of making dynamic predictions. According to the United Nations Development Programme (UNDP) [2], traditional statistical methods may require two years of data collection and analysis to predict poverty. Machine learning will be a great way to empower government and business sectors to make more intelligent and strategic decisions, ultimately supporting the lives of vulnerable people in society and leading towards a sustainable future.

To build a machine learning model for poverty prediction, this study referenced a research paper titled "Poverty Classification Using Machine Learning: The Case of Jordan," which presents a machine learning model to predict poverty among Jordanian households [3]. Alsharkawi *et al*. [3]

implemented a classification model that is robust enough to deal with changes in political, social, and economical factors. This study achieved an F1 score of 81.0% using Gradient Boost implemented with Light GBM, which is an acceptable level of accuracy compared to the average F1 score of 87.1% among poverty prediction classification models (i.e., Naïve Bayes, Decision Tree, K-Nearest Neighbors, Logistic Regression, and ID3) in other countries (i.e., Lagangilang, Abra, and Philippines) [4]. This paper aims to build a machine learning model to predict the poverty status of Costa Rican households.

## II.    EXPLORATORY DATA ANALYSIS

In this study, the Inter-American Development Bank's Costa Rican household dataset was used to build a machine learning model for predicting the poverty status of Costa Rican households. The dataset was compiled through a proxy means test that includes questionnaires related to household composition, observable characteristics of the household (e.g., material of roof), and ownership of electronic devices.

As shown in **Appendix - Exhibit 1**, the dataset has 143 columns (i.e., a mix of categorical and numerical variables) and 9,557 rows. Of the 143 variables, the variable titled "Target" is used as a dependent variable to predict poverty status. This variable consists of four classes (i.e., extreme poverty, moderate poverty, vulnerable households, and non-vulnerable households). To examine whether the dataset is imbalanced, univariate analysis is conducted. As shown in **Exhibit 2**, of the 9,557 survey participants, 5,996 were non-vulnerable households. This represents about 62.7% of the total, which indicates that this is the majority class of the dataset.

***EXHIBIT 2.*** *DEPENDENT VARIABLE UNIVARIATE ANALYSIS*

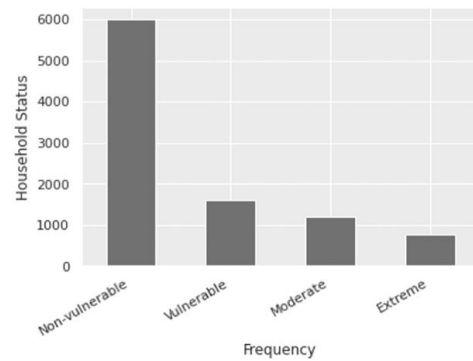

As shown in **Exhibit 3**, 6,829 of the 9,557 survey participants live in urban areas, which represents about 71.4% of the total.

***EXHIBIT 3**. GEOGRAPHICAL REPRESENTATION OF THE SAMPLES*

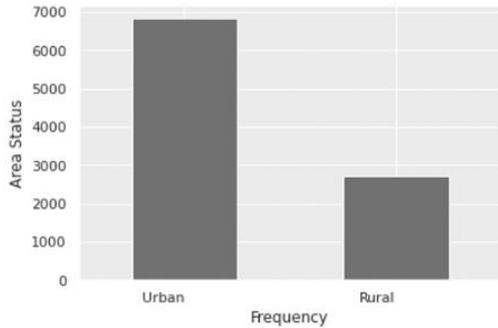

As shown in **Appendix - Exhibit 1**, seven variables (i.e., v2a1, v18q1, dependency, edjefe, edjefa, meaneduc, and SQBmeaned) have missing values. In particular, v2a1 and v18q1 have estimated missing values percentages of 72.0% and 75.0%, respectively.

## III.     APPROACH

In this study, five models are considered (i.e., Decision Tree, Random Forest, Gradient Boosted Trees, Naïve Bayes, and K-Nearest Neighbors).

### A. Decision Tree

***EXHIBIT 4**. DECISION TREE STRUCTURE*

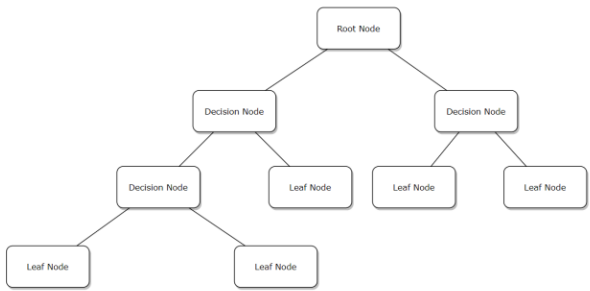

A Decision Tree is a model that utilizes the tree-like model for analyzing and forecasting the data. The tree consists of the root node, internal nodes, and leaf nodes, and is recursively split into sub-trees [5]. A Decision Tree is one of the most widely used machine learning models because the model can handle categorical and numerical datasets, as well as a mix of categorical and numerical datasets. It can also be applied by non-expert users more easily than other machine learning models, as it requires less skill in data pre-processing, and because the model has a built-in resistance to outliers [6]. Decision Trees can be utilized for datasets with missing values; many studies have found the method to work with such datasets [7], [8]. However, because Decision Trees require careful parameter tuning to prevent the model from becoming biased towards the majority class [9], they were not used in this study.

### B. Random Forest

***EXHIBIT 5**. RANDOM FOREST STRUCTURE [10]*

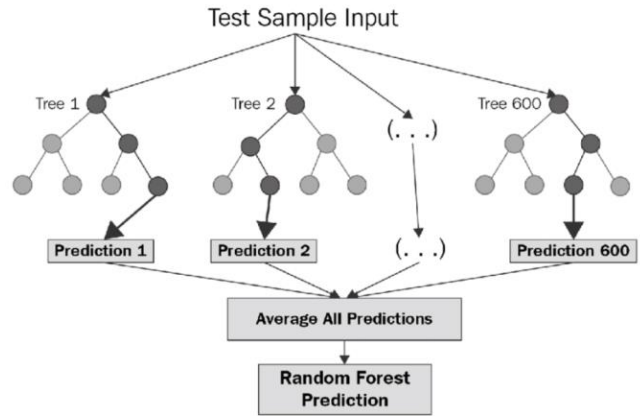

Random Forest creates multiple independent trees using a random sample of data and aggregates trees that are created using a Decision Tree model. By aggregating the results of different trees into one result, Random Forest can limit overfitting without increasing errors that are caused by bias [11]. As Decision Trees can be utilized to overcome missing values, Random Forest is also a well-known algorithm that can handle datasets with missing values. Because Random Forest can decrease the risk of overfitting [12], and because it works well with non-linear data [13], it was used in this study to predict the poverty status of Costa Rican households.

### C. Gradient Boosted Trees

***EXHIBIT 6**. GRADIENT BOOSTED TREES STRUCTURE [14]*

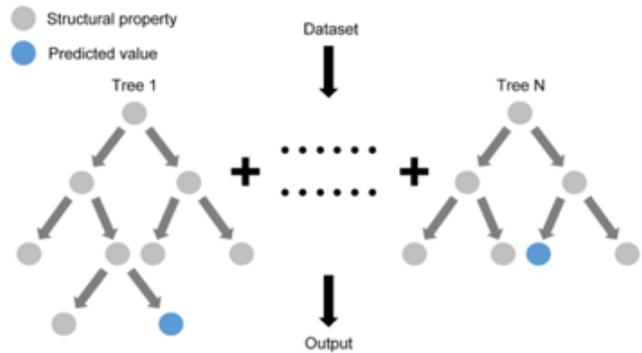

Gradient Boosted Trees can be used to improve the predictive performance of a Decision Tree. Gradient Boosted Trees generate the trees sequentially, and new trees correct previously trained trees iteratively [15], [16]. Gradient Boosted Trees are prone to overfitting, as they develop the models based on the previous trees. However, regularization parameters (e.g., learning rate or shrinkage parameter) prevent overfitting by controlling the amount of information coming from previously fitted trees when forming new trees [17]. Various algorithms can be applied to Gradient Boosted Trees to handle the missing values in the datasets, allowing Gradient Boosted Trees to minimize loss functions and the risks of under/overfitting [18]. Because Gradient Boosted Trees' ability in "minimizing some loss function" makes it "to be more accurate than some more theoretically intensive predictive models" [17, p. 9], Gradient Boosted Trees were used in this study.

### D. Naïve Bayes

**EXHIBIT 7.** *NAÏVE BAYES STRUCTURE*

$$P(A|B_1 \cap B_2 \cap \cdots \cap B_n) = \frac{P(B_1|A)P(B_2|A)\cdots P(B_n|A)P(A)}{P(B)}$$

The Naïve Bayes is a popular algorithm in machine learning because it can be used with large datasets efficiently, and it can be interpreted easily. However, as the word *naïve* suggests, Naïve Bayes assumes that the features are independent [7]. Also, special consideration is needed when using Naïve Bayes with datasets that have both numerical and categorical variables [19]. Thus, the Naïve Bayes was not used in this study, due to the limitations of the model.

### E. K-Nearest Neighbors

**EXHIBIT 8.** *K-NEAREST NEIGHBORS WITH EUCLIDEAN DISTANCE STRUCTURE*

$$\sqrt{\sum_{i=1}^{n}(x_i - y_i)^2}$$

K-Nearest Neighbors can be implemented simply because it is a non-parametric algorithm. It does not require training steps, as it does not build any models [20]. "Instead an observation is predicted to be the class of that of the largest proportion of the k nearest observations" [21, p. 251]. Because K-Nearest Neighbors is sensitive to outliers [22], it was not used in this study.

## IV.    DATA PRE-PROCESSING

### A. Cleaning and Wrangling the Dataset

As shown in **Appendix - Exhibit 9**, multiple individual variables with similar characteristics are merged into one variable. As a result, 18 new variables are formed, and they are encoded as dummy variables along with other ungrouped binary categorical variables.

The age variable is a continuous data type, with a range from 0 to 100. This variable is divided into six groups (i.e., children, adolescents, young adults, adults, middle-aged adults, and old adults). These ordinal groups are mapped with unique labels to transform them from continuous to categorical data types. Because inaccurate binning can add bias to the dataset, numerical variables with dependent relationships to other variables remain as numerical variables.

The dataset is reorganized according to the following criteria. First, when multiple variables contain the same values under different variable names[1], only one variable remains and the rest of the variables are deleted. Second, when the same property [2] is expressed in two different data types (i.e., categorical and numerical), and when multiple variables are similar to each other[3], the variable containing more meaningful information is retained. Third, variables that contain limited information[4] are removed from the dataset.

Of the seven variables with missing values (i.e., v2a1, v18q1, dependency, edjefe, edjefa, meaneduc, and SQBmeaned), two (i.e., v2a1 and v18q1) were deleted from the dataset. Deleting variables can cause a loss of information and introduce bias into the model [21]; however, the proportion of missing values for both variables was too large to be replaced with statistical values (i.e., mean, median, and mode). The remaining four[5] variables (i.e., dependency, edjefe, edjefa, and meaneduc) were replaced with the median value of the variable. Replacing missing values with the median value is not the most accurate approach, so other techniques (e.g., predicting missing values using algorithms) can be considered in future studies.

As shown in **Appendix - Exhibit 9**, 125 variables were used to build the model after data cleaning and wrangling. Among these variables, 17 (i.e., rooms, r4h1, r4h2, r4h3, r4m1, r4m2, r4m3, r4t1, r4t2, r4t3, escolari, rez_esc, dependency, edjefe, edjefa, meaneduc, and overcrowding) are numerical variables, either discrete or continuous data types. To examine their distribution and skewness, these variables were individually plotted, as shown in **Appendix - Exhibit 10.**

### B. Rescaling the Dataset

Normalization and standardization were performed to rescale the distribution of numerical variables. Standardization was used to rescale numerical variables, except dependency variables. Normalization was performed on dependency variables, as they had lower and upper bounds of 0% and 100%.

**EXHIBIT 11.** *NORMALIZATION EQUATION*

$$x'_i = \frac{x_i - \min(x)}{\max(x) - \min(x)}$$

Normalization rescaled variables to fit within the range of 0 to 1. Max($x$) indicates maximum values, and min($x$) indicates minimum values [21].

---

[1] For example, tamhog, tamviv, Hhsize, hogar_total are variables representing the number of household members.

[2] For example, mobilephone is a categorical variable representing whether the participant has a mobile phone, and Qmobiliephone is a discrete variable representing how many mobile phones the household has.

[3] For example, r4h1 is a variable representing the number of children under the age of 12 years in the household, and hogar_nin is a variable representing the number of children between the ages of 0 and 19 in the household.

[4] For example, Id is a variable representing survey participants' identification number.

[5] SQBmeaned was deleted based on dataset reorganization criteria.

**EXHIBIT 12**. *STANDARDIZATION EQUATION*

$$x'_i = \frac{x_i - \bar{x}}{\sigma}$$

Standardization rescaled variables into a normal distribution, with means of 0 and standard deviations of 1. $\bar{x}$ indicates the mean of the variable, and $\sigma$ indicates the standard deviation of the variable [21].

### C. Reducing the Dimensionality of the Dataset

After normalization and standardization, Principal Component Analysis (PCA) was performed to "reduce the dimensionality (number of variables) of the dataset but retain most of the original variability in the data" [23, p. 5]. As shown in **Exhibit 13**, the amount of explained variance above 60 principal components is very low.

**EXHIBIT 13**. *EXPLAINED VARIANCE RATIO OF COMPONENTS*

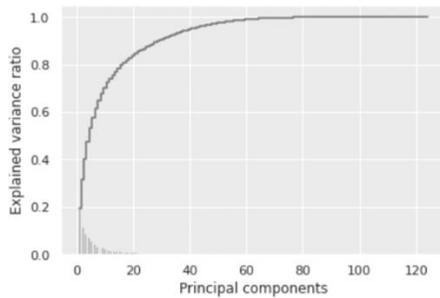

### V. DATA MODELLING

Some important findings were made during exploratory analysis and data cleaning and wrangling. First, the model should be able to deal with the supervised multiclassification problem. Second, the model should be able to work with heterogeneous datasets (i.e., a mix of categorical and numerical variables). Third, the model should excel in processing outliers and missing values. Therefore, Random Forest and Gradient Boosted Trees were selected from the five previously considered models to predict the poverty status of Costa Rican households. Before building models, the study randomly split the dataset into train and test sets. The train set comprises 80% of the dataset, while the test set comprises the other 20%.

### VI. EVALUATION

Because the dataset is imbalanced, stratified five-fold cross-validation was performed on the train set to determine the generalized performance of the model. Stratified cross-validation helps ensure "that the proportions between classes are the same in each fold" [24, p. 255]. The accuracy of the Random Forest model is 76.0%, while the accuracy of the Gradient Boosted Trees model is 77.6%.

To determine the models' predictive power on the test set, accuracy was calculated. The accuracy of Random Forest and Gradient Boosted Trees is 78.1% and 79.6%, respectively.

Because the models are built on an imbalanced dataset, other performance evaluation methods (i.e., F1, recall, and precision) were used to compare the performance of the model under different metrics. As shown in **Exhibit 14**, the Random Forest model achieved the highest score with 88.5%, followed by Gradient Boosted Trees with 82.4%. Random Forest performed well in the precision method, while it underperformed Gradient Boosted Trees in the recall method. Therefore, F1 was used, as this measure provides "the harmonic mean of precision and recall" [4, p. 13]. Gradient Boosted Trees achieved an F1 score of 68.4%, and Random Forest achieved a score of 64.9%.

**EXHIBIT 14**. *MULTICLASSIFICATION MODELS PERFORMANCE RESULTS*

| Model | Accuracy | F1 | Recall | Precision |
|---|---|---|---|---|
| Random Forest | 78.1 | 64.9 | 56.9 | 88.5 |
| Gradient Boosted Trees | 79.6 | 68.4 | 61.7 | 82.4 |

Performance evaluation metrics do not yield the same result as data balancing. Therefore, the most accurate approach will involve equally weighting all four classes through data-balancing techniques (e.g., over-sampling, under-sampling, the synthetic minority over-sampling technique (SMOTE), and class weights) during the data pre-processing. These techniques can be explored further in future studies.

### VII. FEATURE IMPORTANCE

Because Gradient Boosted Trees achieved a higher F1 score than Random Forest, feature importance analysis was performed on Gradient Boosted Trees to determine which variables had the greatest effect on the model. As shown in **Exhibit 15**, the meaneduc variable (i.e., average years of education for adults) was found to be the most impactful variable.

**EXHIBIT 15**. *TOP THREE IMPORTANT COMPONENTS*

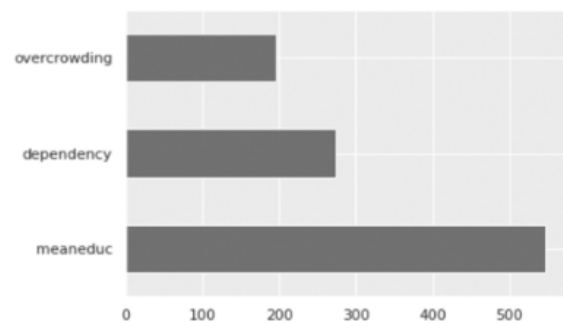

### VIII. LIMITATIONS

This study has several limitations. First, although the dataset was assembled by a credible organization, the Inter-American Development Bank, some information (e.g., the data collection period and dataset creation time) is unavailable; therefore, it is difficult to understand what kinds of variance are included in the dataset.

Second, Costa Rica's population in 2020 was 5,094,114 [25], but the size of the dataset used in this study is 9,557. The small sample size suggests that the population's characteristics may not be adequately represented in the dataset. However, the historical patterns of poverty among Costa Rican households and the percentage of urban populations agree with the dataset. As this dataset was originally published by the Inter-American Development Bank to develop a machine learning model to predict poverty status, this study assumes that collected samples truly reflect the population of Costa Rica. If the collected sample does not reflect the population demographics for some reason (e.g., the sample is collected from the specific regions, or the sample is collected from the specific target), the research findings would less closely reflect the population.

Along with the limitations of the dataset itself, due to resource and time constraints, several important techniques could not be performed. In future studies, these approaches can be applied to improve poverty status prediction.

### A. Handling Missing Values

In this study, two variables (i.e., v2a1 and v18q1) with a significant amount of missing values were deleted, and the missing values of four variables (i.e., dependency, edjefe, edjefa, and meaneduc) were replaced with the median value of those variables. However, these methods are not the most accurate techniques for handling missing values. Because inaccurate data cleaning and wrangling techniques can introduce bias or reduce variance in the dataset, it is important to pre-distinguish the types of missing values (e.g., missing completely at random, missing at random, and missing not at random). Another approach involves predicting the approximate value of missing values using algorithms (e.g., K-Nearest Neighbors). These more precise techniques will correct the reduction in accuracy caused by mishandling missing values.

### B. Handling Dataset Imbalance

As mentioned earlier, among the four classes, the non-vulnerable class comprises approximately 62.7% of the dataset. This indicates that the dataset is imbalanced. Therefore, the dataset has to be balanced to prevent the machine learning model from becoming biased towards the majority class. There are four methods to balance the dataset. The first method is random under-sampling. Random under-sampling will "randomly delete examples in the majority class" [26, p. 113]. The disadvantage of random under-sampling is that "this method can discard potentially useful data that could be important for the induction process" [27, p. 2]. The second method is random over-sampling. Random over-sampling will "randomly duplicate examples in the minority class" [26, p. 113]. Random over-sampling has its own disadvantages as well. Random over-sampling does not "add any new information" to the model, as it involves "duplicating examples in the minority class" [26, p. 121]. To overcome the disadvantages of random over-sampling, SMOTE was invented by researchers. SMOTE "synthesizes new examples for the minority class" [26, p. 121]. However, SMOTE can potentially add noise to the model, as synthetic minority examples are formed with different minority class examples [26]. Lastly, class weights can be used to equally weigh all four data classes. This technique places different weights on each class to emphasize the minority class [3]. All four techniques have their advantages and disadvantages; therefore, future studies can apply these techniques to the model to find the best performing data-balancing method.

### C. Tuning Model Performance

As shown in **Exhibit 14**, performance varies between the four evaluation metrics. Recall and F1 underperform training accuracy, while accuracy and precision outperform training accuracy. Because the dataset is imbalanced, not all classes may be classified equally. The other possibility is that the test set may represent a localized portion of the train set, as it comprises only 20% of the dataset. However, these are just two of many possible explanations for its performance. In future studies, further examination (e.g., building separate one-versus-rest classifiers to review the performance of each class) can be conducted to clearly distinguish factors that may cause under/overfitting and to determine the generalized performance of the model.

### D. Implementing the Naïve Bayes

Other studies have proven that the Naïve Bayes works well in predicting poverty status [4]; therefore, future studies can consider implementing the Naïve Bayes. However, the Naïve Bayes performs well only when variables are independent. Social datasets contain variables that are sometimes highly correlated with each other, forming a dependent relationship. In the future studies, further feature engineering can be attempted to eliminate dependency between variables.

Assuming that independence can be established by eliminating dependency, two approaches can be considered in future studies for implementing the Naïve Bayes in predicting poverty status.

First, binning can be considered as a way to transform a heterogeneous dataset into a homogeneous dataset. Numerical variables can be binned to remove numeric attributes and transform them into categorical variables. However, variables grouped by unspecific criteria can introduce bias to the dataset; thus, binning can be conducted only when specific, objective, and clear criteria is available.

Second, if the dataset cannot be made homogenous, having a mix of categorical and continuous variables, special consideration is needed when implementing the Naïve Bayes classifier. Hsu *et al.* [19] developed the Extended Naïve Bayes (ENB) classifier, in which probabilities of categorical variables are calculated using the original method in the Naïve Bayes model, and variances of numerical variables are found using the statistical theory.

### IX.      CONCLUSION

In this study, a dataset collected through a proxy means test by the Inter-American Development Bank was used to predict the poverty status of Costa Rican households. Based on characteristics of the dataset (i.e., multiclassification, heterogeneous dataset, missing values, and outliers), Random Forest and Gradient Boosted Trees were selected to develop multiclassification poverty prediction models.

Before building Random Forest and Gradient Boosted Trees models, irrelevant or highly correlated variables were deleted, and missing values were replaced with the median value of the variable to simplify the dataset. Both normalization and standardization were used to rescale categorical and numerical variables. Also, PCA was performed to reduce the dimensionality of the dataset.

As a result, under the assumption that the dataset reflects the characteristics of Costa Rica's population, the Random Forest model achieved a 64.9% F1 score, while the Gradient Boosted Trees model achieved a score of 68.4%. However, in terms of F1 scores, these models underperformed the Jordanian model and the average of other models found in the literature.

Further, this study found that education (i.e., meaneduc) has the greatest impact on predicting the status. Finding a causal relationship between educational attainment and poverty was not a goal of this study, so further examination of this topic was not carried out. However, many prominent scholars have revealed that additional years of education increase individual income [28].

Despite their several limitations, both Random Forest and Gradient Boosted Trees demonstrated the ability to predict poverty status among Costa Rican households. Future studies could address the limitations described in this study to improve the performance of these models. Further, the models' robustness could be measured by adding a variety of social and economic factors into the dataset. Such efforts will continue after this study to strengthen the models, as this is an area of research with development potential.

## X.    APPENDIX

*A. **EXHIBIT** 1. ORIGINAL DATASET WITH DESCRIPTIONS OF VARIABLES*

| # | Variable Name | Missing Values | Variable Description |
|---|---|---|---|
| 1 | Id | 0 | Survey participant ID |
| 2 | v2a1 | 6,860 | Monthly rent payment |
| 3 | hacdor | 0 | =1 overcrowding by bedrooms |
| 4 | Rooms | 0 | # of all rooms in the house |
| 5 | hacapo | 0 | =1 overcrowding by rooms |
| 6 | v14a | 0 | =1 if the household has a toilet |
| 7 | Refrig | 0 | =1 if the household has a refrigerator |
| 8 | v18q | 0 | =1 if the household has a tablet |
| 9 | v18q1 | 7,342 | # of tablets household owns |
| 10 | r4h1 | 0 | # of males younger than 12 years of age |
| 11 | r4h2 | 0 | # of males 12 years of age and older |
| 12 | r4h3 | 0 | # of males in the household |
| 13 | r4m1 | 0 | # of females younger than 12 years of age |
| 14 | r4m2 | 0 | # of females 12 years of age and older |
| 15 | r4m3 | 0 | # of females in the household |
| 16 | r4t1 | 0 | # of persons younger than 12 years of age |
| 17 | r4t2 | 0 | # of persons 12 years of age and older |
| 18 | r4t3 | 0 | # of persons in the household |
| 19 | tamhog | 0 | Household size |
| 20 | tamviv | 0 | Household size |
| 21 | escolari | 0 | # of years of schooling |
| 22 | rez_esc | 0 | # of years behind in school |
| 23 | Hhsize | 0 | Household size |
| 24 | paredblolad | 0 | =1 if predominant material on the outside wall is block or brick |
| 25 | paredzocalo | 0 | =1 if predominant material on the outside wall is socket (wood, zinc, or asbestos) |
| 26 | paredpreb | 0 | =1 if predominant material on the outside wall is prefabricated or cement |
| 27 | pareddes | 0 | =1 if predominant material on the outside wall is waste material |
| 28 | paredmad | 0 | =1 if predominant material on the outside wall is wood |
| 29 | paredzinc | 0 | =1 if predominant material on the outside wall is zinc |

| # | Variable Name | Missing Values | Variable Description |
|---|---|---|---|
| 30 | paredfibras | 0 | =1 if predominant material on the outside wall is natural material |
| 31 | paredother | 0 | =1 if predominant material on the outside wall is other |
| 32 | pisomoscer | 0 | =1 if predominant material on the floor is mosaic, ceramic, or terrazzo |
| 33 | pisocemento | 0 | =1 if predominant material on the floor is cement |
| 34 | pisoother | 0 | =1 if predominant material on the floor is other |
| 35 | pisonatur | 0 | =1 if predominant material on the floor is natural material |
| 36 | pisonotiene | 0 | =1 if no floor at the household |
| 37 | pisomadera | 0 | =1 if predominant material on the floor is wood |
| 38 | techozinc | 0 | =1 if predominant material on the roof is metal foil or zinc |
| 39 | techoentrepiso | 0 | =1 if predominant material on the roof is fiber cement or mezzanine |
| 40 | techocane | 0 | =1 if predominant material on the roof is natural material |
| 41 | techootro | 0 | =1 if predominant material on the roof is other |
| 42 | cielorazo | 0 | =1 if the house has a ceiling |
| 43 | abastaguadentro | 0 | =1 if water provision inside the dwelling |
| 44 | abastaguafuera | 0 | =1 if water provision outside the dwelling |
| 45 | abastaguano | 0 | =1 if no water provision |
| 46 | Public | 0 | =1 electricity from CNFL, ICE, or ESPH/JASEC |
| 47 | planpri | 0 | =1 electricity from private plant |
| 48 | noelec | 0 | =1 no electricity in the dwelling |
| 49 | coopele | 0 | =1 electricity from cooperative |
| 50 | sanitario1 | 0 | =1 no toilet in the dwelling |
| 51 | sanitario2 | 0 | =1 toilet connected to sewer or cesspool |
| 52 | sanitario3 | 0 | =1 toilet connected to septic tank |
| 53 | sanitario5 | 0 | =1 toilet connected to hole or latrine |
| 54 | sanitario6 | 0 | =1 toilet connected to other system |
| 55 | energcocinar1 | 0 | =1 no main source of energy used for cooking (no kitchen) |
| 56 | energcocinar2 | 0 | =1 main source of energy used for cooking is electricity |
| 57 | energcocinar3 | 0 | =1 main source of energy used for cooking is gas |
| 58 | energcocinar4 | 0 | =1 main source of energy used for cooking is wood charcoal |
| 59 | elimbasu1 | 0 | =1 if rubbish is disposed mainly by tanker truck |
| 60 | elimbasu2 | 0 | =1 if rubbish is disposed mainly by botan hollow or buried |
| 61 | elimbasu3 | 0 | =1 if rubbish is disposed mainly by burning |
| 62 | elimbasu4 | 0 | =1 if rubbish is disposed mainly by throwing in an unoccupied space |
| 63 | elimbasu5 | 0 | =1 if rubbish is disposed mainly by throwing in river, creek, or sea |
| 64 | elimbasu6 | 0 | =1 if rubbish is disposed mainly by other |
| 65 | epared1 | 0 | =1 if walls are bad |
| 66 | epared2 | 0 | =1 if walls are regular |
| 67 | epared3 | 0 | =1 if walls are good |
| 68 | etecho1 | 0 | =1 if roof is bad |
| 69 | etecho2 | 0 | =1 if roof is regular |
| 70 | etecho3 | 0 | =1 if roof is good |
| 71 | eviv1 | 0 | =1 if floor is bad |
| 72 | eviv2 | 0 | =1 if floor is regular |
| 73 | eviv3 | 0 | =1 if floor is good |
| 74 | Dis | 0 | =1 if disabled person |
| 75 | Male | 0 | =1 if male |
| 76 | female | 0 | =1 if female |
| 77 | estadocivil1 | 0 | =1 if less than 10 years old |
| 78 | estadocivil2 | 0 | =1 if free or coupled union |
| 79 | estadocivil3 | 0 | =1 if married |
| 80 | estadocivil4 | 0 | =1 if divorced |
| 81 | estadocivil5 | 0 | =1 if separated |
| 82 | estadocivil6 | 0 | =1 if widow/er |
| 83 | estadocivil7 | 0 | =1 if single |

| # | Variable Name | Missing Values | Variable Description |
|---|---|---|---|
| 84 | parentesco1 | 0 | =1 if household head |
| 85 | parentesco2 | 0 | =1 if spouse/partner |
| 86 | parentesco3 | 0 | =1 if son/daughter |
| 87 | parentesco4 | 0 | =1 if stepson/daughter |
| 88 | parentesco5 | 0 | =1 if son/daughter-in-law |
| 89 | parentesco6 | 0 | =1 if grandson/daughter |
| 90 | parentesco7 | 0 | =1 if mother/father |
| 91 | parentesco8 | 0 | =1 if father/mother-in-law |
| 92 | parentesco9 | 0 | =1 if brother/sister |
| 93 | parentesco10 | 0 | =1 if brother/sister-in-law |
| 94 | parentesco11 | 0 | =1 if other family member |
| 95 | parentesco12 | 0 | =1 if other non-family member |
| 96 | idhogar | 0 | Household level identifier |
| 97 | hogar_nin | 0 | # of children 0 to 19 in household |
| 98 | hogar_adul | 0 | # of adults in household |
| 99 | hogar_mayor | 0 | # of individuals 65+ in the household |
| 100 | hogar_total | 0 | # of total individuals in the household |
| 101 | dependency | 2,192 | Dependency rate |
| 102 | Edjefe | 123 | # of years of education of male head of household |
| 103 | Edjefa | 69 | # of years of education of female head of household |
| 104 | meaneduc | 5 | Average years of education for adults (18+) |
| 105 | instlevel1 | 0 | =1 no level of education |
| 106 | instlevel2 | 0 | =1 incomplete primary |
| 107 | instlevel3 | 0 | =1 complete primary |
| 108 | instlevel4 | 0 | =1 incomplete academic secondary level |
| 109 | instlevel5 | 0 | =1 complete academic secondary level |
| 110 | instlevel6 | 0 | =1 incomplete technical secondary level |
| 111 | instlevel7 | 0 | =1 complete technical secondary level |
| 112 | instlevel8 | 0 | =1 undergraduate and higher education |
| 113 | instlevel9 | 0 | =1 postgraduate higher education |
| 114 | bedrooms | 0 | # of bedrooms |
| 115 | overcrowding | 0 | Persons per room |
| 116 | tipovivi1 | 0 | =1 own and fully paid house |
| 117 | tipovivi2 | 0 | =1 own, paying in installments |
| 118 | tipovivi3 | 0 | =1 rented |
| 119 | tipovivi4 | 0 | =1 precarious |
| 120 | tipovivi5 | 0 | =1 other(assigned or borrowed) |
| 121 | computer | 0 | =1 if the household has a notebook or desktop computer |
| 122 | television | 0 | =1 if the household has a TV |
| 123 | mobilephone | 0 | =1 if the household has a mobile phone |
| 124 | qmobilephone | 0 | # of mobile phones household owns |
| 125 | lugar1 | 0 | =1 region Central |
| 126 | lugar2 | 0 | =1 region Chorotega |
| 127 | lugar3 | 0 | =1 region PacÃƒÂ­fico[6] central |
| 128 | lugar4 | 0 | =1 region Brunca |
| 129 | lugar5 | 0 | =1 region Huetar AtlÃƒÂ¡ntica[7] |
| 130 | lugar6 | 0 | =1 region Huetar Norte |
| 131 | area1 | 0 | =1 zona urbana |
| 132 | area2 | 0 | =2 zona rural |
| 133 | Age | 0 | Age in years |
| 134 | SQBescolari | 0 | Escolari squared |

---

[6] The original dataset has a character encoding error.
[7] The original dataset has a character encoding error.

| # | Variable Name | Missing Values | Variable Description |
|---|---|---|---|
| 135 | SQBage | 0 | Age squared |
| 136 | SQBhogar_total | 0 | Hogar_total squared |
| 137 | SQBedjefe | 0 | Edjefe squared |
| 138 | SQBhogar_nin | 0 | Hogar_nin squared |
| 139 | SQBovercrowding | 0 | Overcrowding squared |
| 140 | SQBdependency | 0 | Dependency squared |
| 141 | SQBmeaned | 5 | Meaneduc squared |
| 142 | Agesq | 0 | Age squared |
| 143 | Target | 0 | Poverty level |

*B. EXHIBIT 9. DATASET AFTER PRE-PROCESSING*

| # | Variable Name | Variable Type | Variable Description |
|---|---|---|---|
| 1 | Rooms | Discrete | # of all rooms in the house |
| 2 | Refrig | Categorical | =1 if the household has a refrigerator |
| 3 | v18q | Categorical | =1 if the household has a tablet |
| 4 | r4h1 | Discrete | # of males younger than 12 years of age |
| 5 | r4h2 | Discrete | # of males 12 years of age and older |
| 6 | r4h3 | Discrete | #of males in the household |
| 7 | r4m1 | Discrete | # of females younger than 12 years of age |
| 8 | r4m2 | Discrete | # of females 12 years of age and older |
| 9 | r4m3 | Discrete | # of females in the household |
| 10 | r4t1 | Discrete | # of persons younger than 12 years of age |
| 11 | r4t2 | Discrete | # of persons 12 years of age and older |
| 12 | r4t3 | Discrete | # of persons in the household |
| 13 | escolari | Discrete | # of years of schooling |
| 14 | rez_esc | Discrete | # of years behind in school |
| 15 | paredblolad | Categorical | =1 if predominant material on the outside wall is block or brick |
| | paredzocalo | | =1 if predominant material on the outside wall is socket (wood, zinc, or asbestos) |
| | paredpreb | | =1 if predominant material on the outside wall is prefabricated or cement |
| | pareddes | | =1 if predominant material on the outside wall is waste material |
| | paredmad | | =1 if predominant material on the outside wall is wood |
| | paredzinc | | =1 if predominant material on the outside wall is zinc |
| | paredfibras | | =1 if predominant material on the outside wall is natural material |
| | paredother | | =1 if predominant material on the outside wall is other |
| 16 | pisomoscer | Categorical | =1 if predominant material on the floor is mosaic, ceramic, or terrazzo |
| | pisocemento | | =1 if predominant material on the floor is cement |
| | pisoother | | =1 if predominant material on the floor is other |
| | pisonatur | | =1 if predominant material on the floor is natural material |
| | pisonotiene | | =1 if no floor at the household |
| | pisomadera | | =1 if predominant material on the floor is wood |
| 17[8] | techozinc | Categorical | =1 if predominant material on the roof is metal foil or zinc |
| | techoentrepiso | | =1 if predominant material on the roof is fiber cement, or mezzanine |
| | techocane | | =1 if predominant material on the roof is natural material |
| | techootro | | =1 if predominant material on the roof is other |
| 18 | cielorazo | Categorical | =1 if the house has a ceiling |
| 19 | abastaguadentro | Categorical | =1 if water provision inside the dwelling |
| | abastaguafuera | | =1 if water provision outside the dwelling |
| | abastaguano | | =1 if no water provision |
| 20[9] | Public | Categorical | =1 electricity from CNFL, ICE, or ESPH/JASEC |
| | planpri | | =1 electricity from private plant |

---

[8] Variables in this row are grouped together based on a characteristic (i.e., roof materials) to encode to dummy variables; however, the study determined that 66 observations are not applicable to any of the four variables in the group. These observations were deleted in this study.

[9] Variables in this row are grouped together based on a characteristic (i.e., electricity type) to encode to dummy variables; however, the study determined that 15 observations are not applicable to any of the four variables in the group. These observations were deleted in this study.

| # | Variable Name | Variable Type | Variable Description |
|---|---|---|---|
| | noelec | | =1 no electricity in the dwelling |
| | coopele | | =1 electricity from cooperative |
| 21 | sanitario1 | | =1 no toilet in the dwelling |
| | sanitario2 | | =1 toilet connected to sewer or cesspool |
| | sanitario3 | Categorical | =1 toilet connected to septic tank |
| | sanitario5 | | =1 toilet connected to hole or latrine |
| | sanitario6 | | =1 toilet connected to other system |
| 22 | energcocinar1 | | =1 no main source of energy used for cooking (no kitchen) |
| | energcocinar2 | | =1 main source of energy used for cooking is electricity |
| | energcocinar3 | Categorical | =1 main source of energy used for cooking is gas |
| | energcocinar4 | | =1 main source of energy used for cooking is wood charcoal |
| 23 | elimbasu1 | | =1 if rubbish is disposed mainly by tanker truck |
| | elimbasu2 | | =1 if rubbish is disposed mainly by botan hollow or buried |
| | elimbasu3 | | =1 if rubbish is disposed mainly by burning |
| | elimbasu4 | Categorical | =1 if rubbish is disposed mainly by throwing in an unoccupied space |
| | elimbasu5 | | =1 if rubbish is disposed mainly by throwing in river, creek, or sea |
| | elimbasu6 | | =1 if rubbish is disposed mainly by other |
| 24 | epared1 | | =1 if walls are bad |
| | epared2 | Categorical | =1 if walls are regular |
| | epared3 | | =1 if walls are good |
| 25 | etecho1 | | =1 if roof is bad |
| | etecho2 | Categorical | =1 if roof is regular |
| | etecho3 | | =1 if roof is good |
| 26 | eviv1 | | =1 if floor is bad |
| | eviv2 | Categorical | =1 if floor is regular |
| | eviv3 | | =1 if floor is good |
| 27 | Dis | Categorical | =1 if disabled person |
| 28 | Male | Categorical | =1 if male |
| | female | | =1 if female |
| 29 | Estadocivil1 | | =1 if less than 10 years old |
| | estadocivil2 | | =1 if free or coupled union |
| | estadocivil3 | | =1 if married |
| | estadocivil4 | Categorical | =1 if divorced |
| | estadocivil5 | | =1 if separated |
| | estadocivil6 | | =1 if widow/er |
| | estadocivil7 | | =1 if single |
| 30 | parentesco1 | | =1 if household head |
| | parentesco2 | | =1 if spouse/partner |
| | parentesco3 | | =1 if son/daughter |
| | parentesco4 | | =1 if stepson/daughter |
| | parentesco5 | | =1 if son/daughter-in-law |
| | parentesco6 | Categorical | =1 if grandson/daughter |
| | parentesco7 | | =1 if mother/father |
| | parentesco8 | | =1 if father/mother-in-law |
| | parentesco9 | | =1 if brother/sister |
| | parentesco10 | | =1 if brother/sister-in-law |
| | parentesco11 | | =1 if other family member |
| | parentesco12 | | =1 if other non-family member |
| 31 | dependency | Continuous | Dependency rate |
| 32 | Edjefe | Discrete | # of years of education of male head of household |
| 33 | Edjefa | Discrete | # of years of education of female head of household |
| 34 | meaneduc | Continuous | Average years of education for adults (18+) |
| | instlevel1 | | =1 no level of education |
| | instlevel2 | Categorical | =1 incomplete primary |

| # | Variable Name | Variable Type | Variable Description |
|---|---|---|---|
| 35[10] | instlevel3 | | =1 complete primary |
| | instlevel4 | | =1 incomplete academic secondary level |
| | instlevel5 | | =1 complete academic secondary level |
| | instlevel6 | | =1 incomplete technical secondary level |
| | instlevel7 | | =1 complete technical secondary level |
| | instlevel8 | | =1 undergraduate and higher education |
| | instlevel9 | | =1 postgraduate higher education |
| 36 | overcrowding | Continuous | Persons per room |
| 37 | tipovivi1 | | =1 own and fully paid house |
| | tipovivi2 | | =1 own, paying in installments |
| | tipovivi3 | Categorical | =1 rented |
| | tipovivi4 | | =1 precarious |
| | tipovivi5 | | =1 other(assigned or borrowed) |
| 38 | computer | Categorical | =1 if the household has a notebook or desktop computer |
| 39 | television | Categorical | =1 if the household has a TV |
| 40 | mobilephone | Categorical | =1 if the household has a mobile phone |
| 41 | lugar1 | | =1 region Central |
| | lugar2 | | =1 region Chorotega |
| | lugar3 | Categorical | =1 region PacÃƒÂfico[11] central |
| | lugar4 | | =1 region Brunca |
| | lugar5 | | =1 region Huetar AtlÃƒÂ¡ntica[12] |
| | lugar6 | | =1 region Huetar Norte |
| 42 | area1 | Categorical | =1 zona urbana |
| | area2 | | =2 zona rural |
| 43 | Age | Categorical | Age in years |
| 44 | Target | Dependent | Poverty level |

*C. **EXHIBIT 10**. DISTRIBUTION OF NUMERICAL VARIABLES*

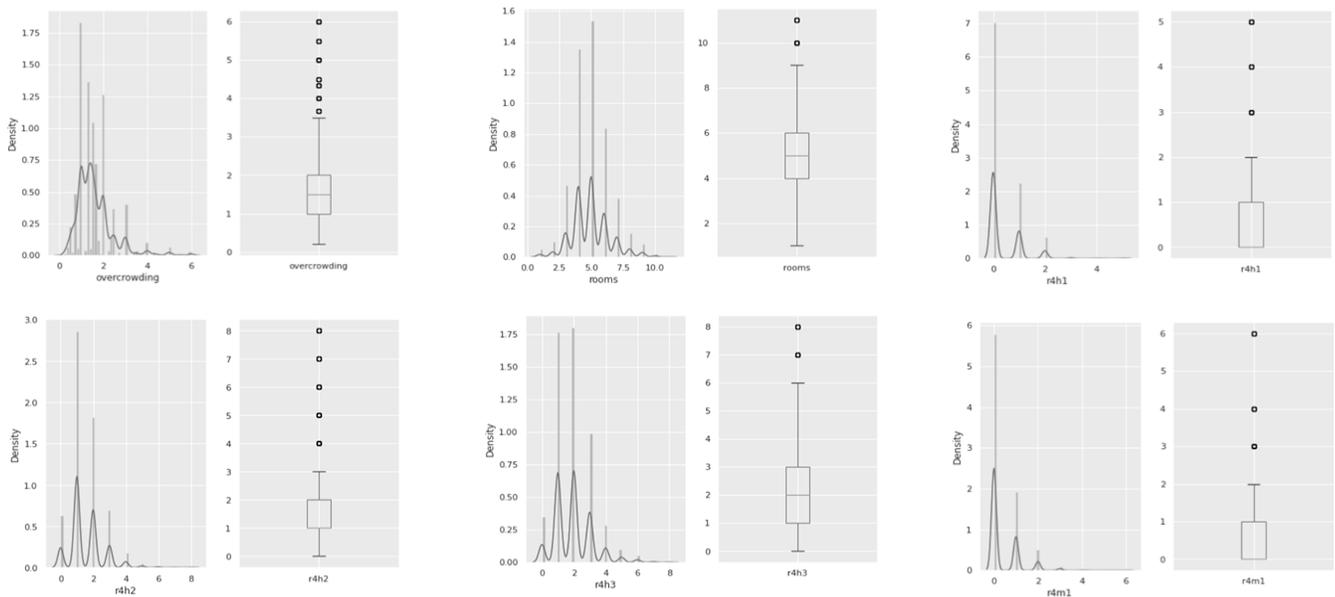

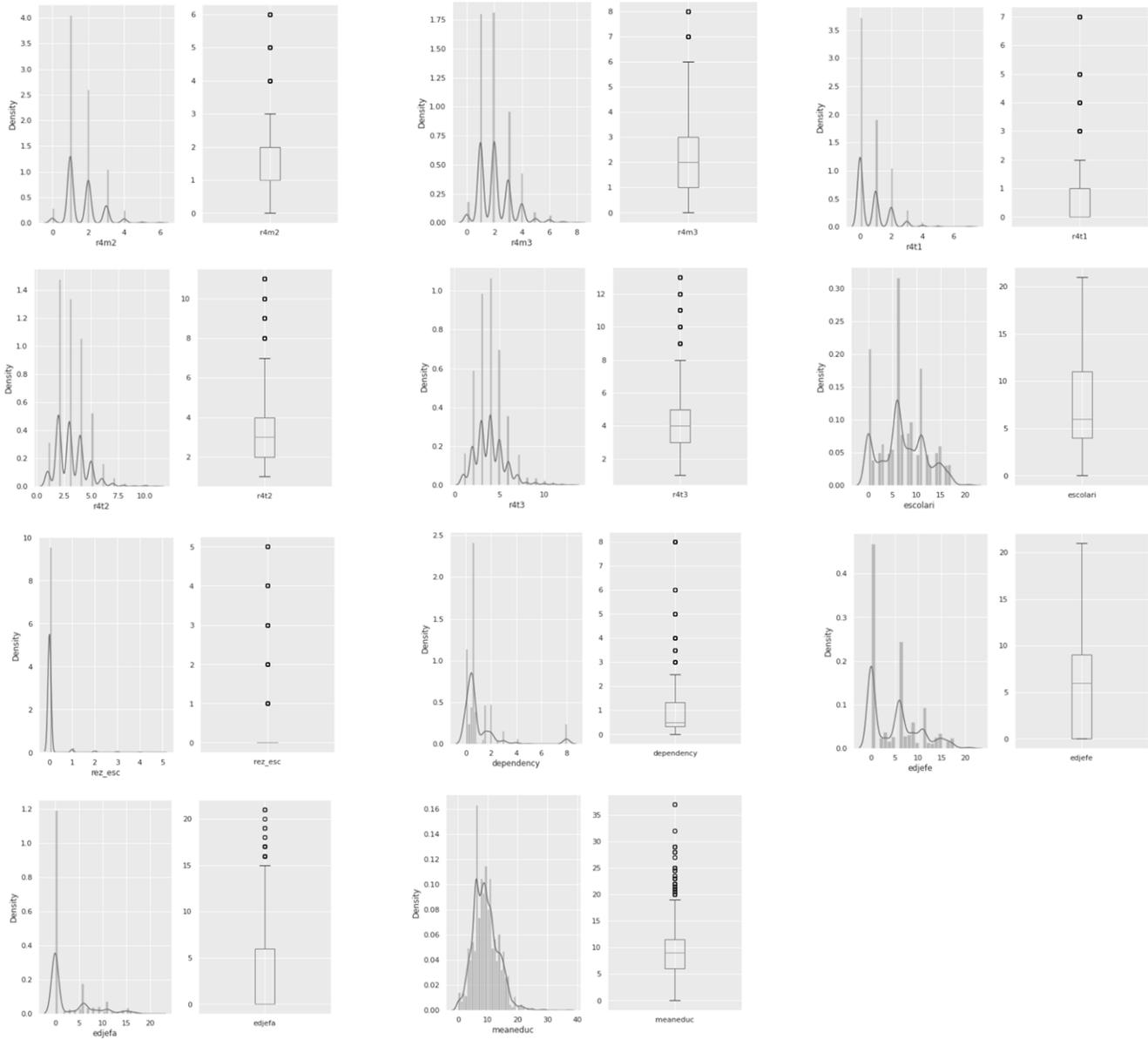